\documentclass[aps,prd,secnumarabic,nobibnotes,twocolumn,superscriptaddress]{revtex4-1}
\usepackage{amsfonts}
\usepackage{mathrsfs}
\usepackage{amsmath}
\usepackage{color}
\usepackage{natbib}
\usepackage{graphicx}
\usepackage{bm}
\usepackage{amssymb}
\usepackage{xspace}
\usepackage{epstopdf}
\usepackage{dcolumn}
\usepackage{multirow}
\usepackage[colorlinks=true, letterpaper=true, pdfstartview=FitV, linkcolor=blue, citecolor=blue, urlcolor=blue]{hyperref}
\usepackage{wrapfig}

\makeatletter

\newcommand{\Rmnum}[1]{\expandafter\@slowromancap\romannumeral #1@}
\makeatother

\begin{document}
\title{Coexistence of zero-, one- and two-dimensional degeneracy in tetragonal SnO$_2$ phonons}

\author{Jianhua Wang}\thanks{J.W. and H.Y. contributed equally to this manuscript.}
\address{School of Physical Science and Technology, Southwest University, Chongqing 400715, China}

\author{Hongkuan Yuan}\thanks{J.W. and H.Y. contributed equally to this manuscript.}
\address{School of Physical Science and Technology, Southwest University, Chongqing 400715, China}

\author{Minquan Kuang}
\address{School of Physical Science and Technology, Southwest University, Chongqing 400715, China}

\author{Tie Yang}
\address{School of Physical Science and Technology, Southwest University, Chongqing 400715, China}

\author{Zhi-Ming Yu}
\address{Centre for Quantum Physics, Key Laboratory of Advanced Optoelectronic Quantum Architecture and Measurement (MOE), School of Physics, Beijing Institute of Technology, Beijing, 100081, China}
\address{Beijing Key Lab of Nanophotonics $\&$ Ultrafine Optoelectronic Systems, School of Physics, Beijing Institute of Technology, Beijing, 100081, China}

\author{Zeying Zhang}\thanks{Corresponding authors}\email{zzy@mail.buct.edu.cn}
\address{College of Mathematics and Physics, Beijing University of Chemical Technology, Beijing 100029, China}

\author{Xiaotian Wang}\thanks{Corresponding authors}\email{xiaotianwang@swu.edu.cn}
\address{School of Physical Science and Technology, Southwest University, Chongqing 400715, China}

\begin{abstract}
Based on the dimension of degeneracy, topological electronic systems can roughly be divided into three parts: nodal point, line and surface materials corresponding to zero-, one- and two-dimensional degeneracy, respectively. In parallel to electronic systems, the concept of topology was extended to phonons, promoting the birth of topological phonons. Till date, few nodal point, line and surface phonons candidates have been predicted in solid-state materials. In this study, based on symmetry analysis and first-principles calculation, for the first time, we prove that zero-, one- and two-dimensional degeneracy co-exist in the phonon dispersion of one single realistic solid-state material SnO$_2$ with \textit{P}4$_2$/\textit{mnm} structure. In contrast to the previously reported electronic systems, the topological phonons observed in SnO$_2$ are not restricted by the Pauli exclusion principle, and they experience negligible spin-orbit coupling effect. Hence, SnO$_2$ with multiple dimensions of degeneracy phonons is a good platform for studying the entanglement among nodal point, line and surface phonons. Moreover, obvious phonon surface states are visible, which is beneficial for experimental detection.
\end{abstract}
\maketitle


The discovery of topological semimetals in electronic systems~\cite{add1,add2,add3,add4,add5} has led to extensive research on the classification of topological states and the identification of materials with topological electronic structures. Generally, topological semimetals obey the following standard of classification~\cite{add6}: (i) dimension of degeneracy (such as nodal point, line or surface); (ii) degree of degeneracy (such as two-, three-, four-, six- or eight-fold degeneracy); and (iii) the type of dispersion (linear, quadratic or cubic). Thus far, thousands of realistic materials~\cite{add7,add8,add9,add10,add11,add12,add13,add14,add15,add16,add17,add18,add19,add20} with topological electronic structures have been proposed using the first-principles calculation. Furthermore, owing to the development of symmetry-based indicator theories~\cite{add18, add21,add22,add23} and compatibility relations~\cite{add24}, they have been widely used to determine the topological state in the electronic structure of most known materials. Moreover, an encyclopedia of quasiparticles in three-dimensional materials has been recently afforded by Yu \textit{et. al.}~\cite{add25} with detailed correspondence among quasiparticles, symmetry conditions, effective models, and topological properties. Despite important achievements in electronic fermionic quasiparticles and symmetry analysis, the study of the topological bosonic state in crystalline materials is still in its infancy. Particularly, the ideal topological semimetal candidates in real electronic materials are very limited due to the strong Fermi level constraint. In contrast, topological bosonic states do not have this constraint and are expected to be great platforms for realizing ideal topological states. However, only a few real materials with topological phonons have been predicted~\cite{add26,add27,add28,add29,add30,add31,add32,add33,add34,add35,add36,add37,add38,add39}, significantly obstructing the development of topological states in phonon systems.

\begin{figure}
\includegraphics[width=8.5cm]{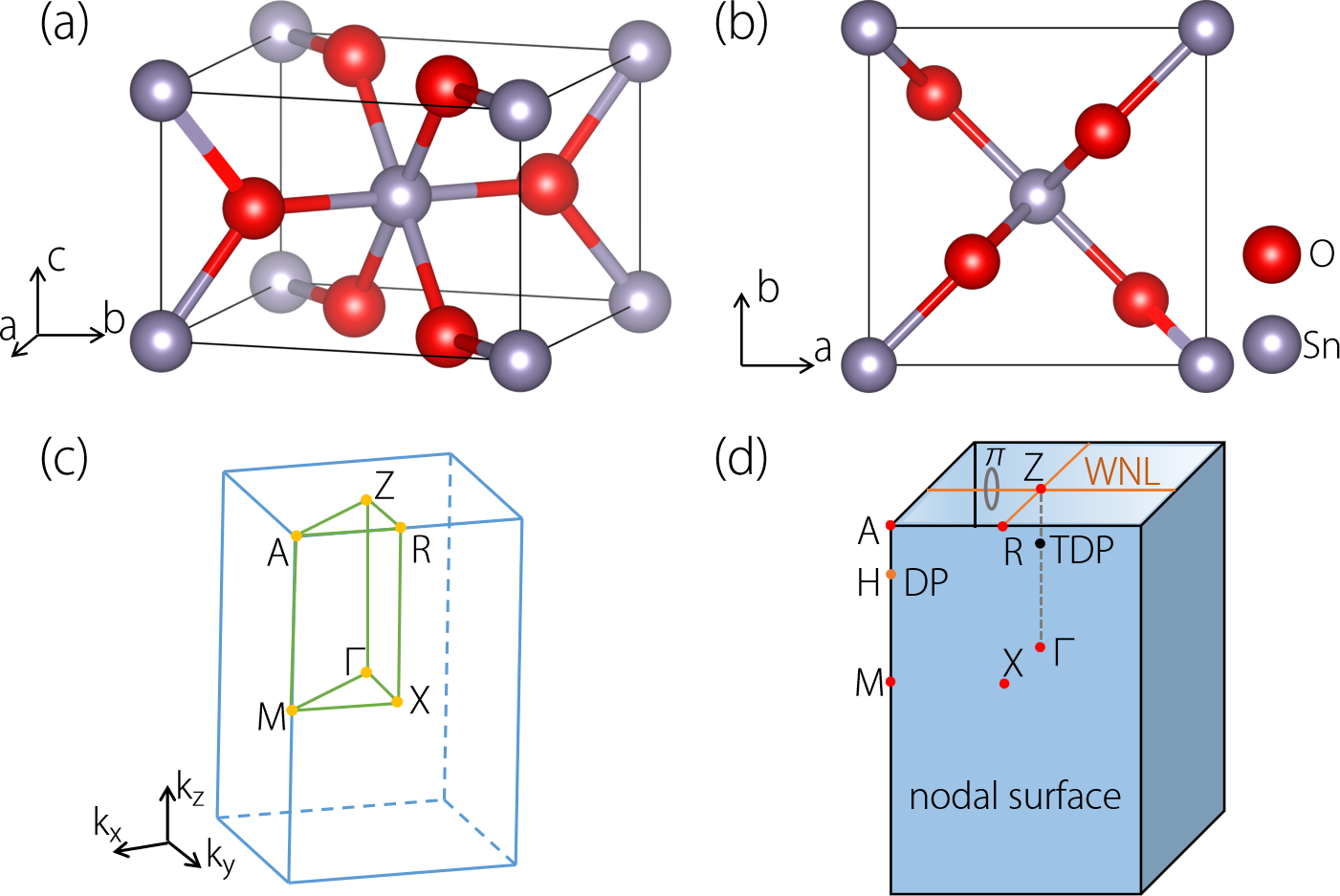}
\caption{(a) and (b) crystal structures of rutile-type SnO$_2$ with \textit{P}4$_2$/\textit{mnm} structure with views of different sides; (c) Three-dimensional coordinate system of BZ and some high symmetry points (A, Z, R, M, $\Gamma$ and X); (d) the schematic of the zero-dimensional Dirac point (DP) and triply degenerate point (TDP), one-dimensional Weyl nodal lines (WNLs), and two-dimensional nodal surface phonons in momentum space. The WNLs possess a quantized $\pi$ Berry phase.
\label{fig1}}
\end{figure}

Topological phonons~\cite{add40,add41,add42,add43,add44,add45} are the basic emergent bosons of crystalline lattices. In parallel to electronic systems, realistic solid-state materials~\cite{add26,add27,add28,add29,add30,add31,add32,add33,add34,add35,add36,add37,add38,add39} have been predicted to comprise nodal point, line and surface phonons with zero-, one- and two-dimensional degeneracy phonons in momentum space, respectively. Moreover, double Weyl point phonons and helical nodal line phonons have been experimentally confirmed in FeSi-type systems and MoB$_2$~\cite{add26,add30}, respectively. However, the above-mentioned studies have only focused on materials with one or two types of dimensions of the degeneracy phonons. \emph{Thus, a natural question is whether three types of dimensions of degeneracy phonons exist in a single synthesized solid-state material.} This study answers this question in the affirmative.

Although three types of fermions (\textit{i.e.}, zero-, one-, two-dimensional degeneracy) have been proposed before in some electronic systems~\cite{add46,add47,add48}, they do not belong to the ideal candidate materials as the proposed topological elements are not near the Fermi level. As discussed above, in contrast to electronic systems, phonons are not limited by the Pauli exclusion principle, and thus, the entire frequency range of the phonon dispersion can be physically observed and used to search for topological elements. Herein, we focus on the solid-state material SnO$_2$ with \textit{P}4$_2$/\textit{mnm} structure~\cite{add49} to prove the presence of zero-, one- and two-dimensional degeneracy phonons in its phonon dispersion. Furthermore, a systematic symmetry analysis for these proposed phonons was performed to provide a deeper physics understanding. Obvious phonon surface states are visible for SnO$_2$; this will benefit their detection in experiments.

In this study, we use the density functional theory to deal with the ground state of the \textit{P}4$_2$/\textit{mnm}-type SnO$_2$ material. For the exchange-correlation functional, GGA-PBE formalism~\cite{add50} is employed. The projector augmented-wave method~\cite{add51} is used for the interactions between ions and valence electrons, and the energy cutoff is set to 600 eV. A $\Gamma$-centered $k$-mesh of 7 $\times$ 7 $\times$ 10 is used for Brillouin zone (BZ) sampling. A lattice dynamic calculation based on the density functional perturbation theory within the PHONOPY package~\cite{add52} is performed to achieve SnO$_2$ phonon dispersion in its ground state. The [001] and [100] phonon surface states of SnO$_2$ are calculated by constructing a Wannier tight-binding Hamiltonian of phonons~\cite{add53}. A prerequisite for the symmetry analysis of the topological states is to obtain the corresponding irreducible representation (IRR) of the phonon spectra. Fortunately, Liu \textit{et. al.} recently developed a SpaceGroupIrep package~\cite{add54} that can be used to calculate the IRR of the phonon spectra for all high-symmetry momentum points. The SpaceGroupIrep package is based on the notation of Ref.~\cite{add55} and is compatible with Ref.~\cite{add25}. We used this package to obtain the required IRRs in this work.

This study focuses on the already prepared rutile-type SnO$_2$ with \textit{P}4$_2$/\textit{mnm} structure~\cite{add49}; the crystal structure with views of different sides of SnO$_2$ is shown in Fig. ~\ref{fig1}(a) and~\ref{fig1}(b). This material is totally relaxed herein and the obtained lattice constants (a = b = 4.801 {\r{A}} and c = 3.242 {\r{A}}) are in good agreements with the experimental ones (a = b = 4.737 {\r{A}} and c = 3.185 {\r{A}})~\cite{add49}. Note that a series of methods for growing SnO$_2$ single crystals have been previously proposed~\cite{add56}, and we believe that the easily obtained SnO$_2$ single crystal is a good platform to experimentally confirm the topological elements mentioned in the following parts of this letter.

\begin{figure}
\includegraphics[width=8.5cm]{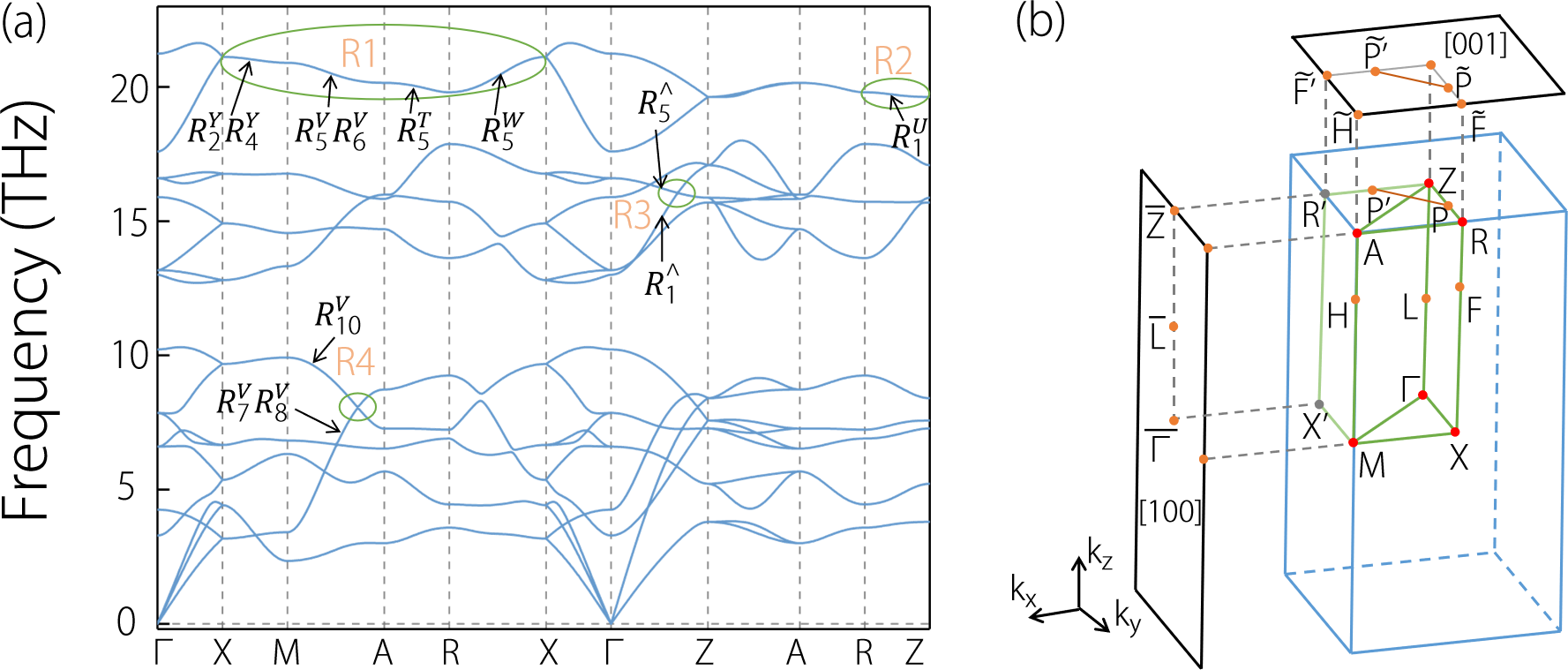}
\caption{(a) The phonon dispersion of SnO$_2$ along the $\Gamma$--X--M--A--R--X--$\Gamma$--Z--A--R--Z path. The two obvious phonon band degenerated regions along X--M--A--R--X and R--Z paths are named as regions R1 and R2, respectively. Two clear phonon band-crossing regions are present along $\Gamma$--Z and M--A regions, which are named as R3 and R4, respectively. The (co)representations for each band in R1-R4 are labeled. (b) The bulk and surface BZ along the [100] and [001] directions of SnO$_2$.
\label{fig2}}
\end{figure}

The SnO$_2$ phonon dispersion along the $\Gamma$--X--M--A--R--X--$\Gamma$--Z--A--R--Z path (see Fig.~\ref{fig1}(c)) is shown in Fig.~\ref{fig2}(a). Obviously, all phonon frequencies are positive, denoting that SnO$_2$ with \textit{P}4$_2$/\textit{mnm} structure is dynamical stable. Careful observation shows that two two-fold degenerate phonon bands exist along the X--M--A--R--X and R--Z paths around 20 THz, named as R1 and R2, respectively. Moreover, two phonon band-crossing points exist: one along $\Gamma$--Z and the other along M--A, named as R3 and R4, respectively. Phonons are bosons, and thus, the entire frequency range of the phonon dispersion can be considered to identify topological elements. Hence, we will separately discuss the topological element in R1-R4.

\begin{figure}
\includegraphics[width=8.5cm]{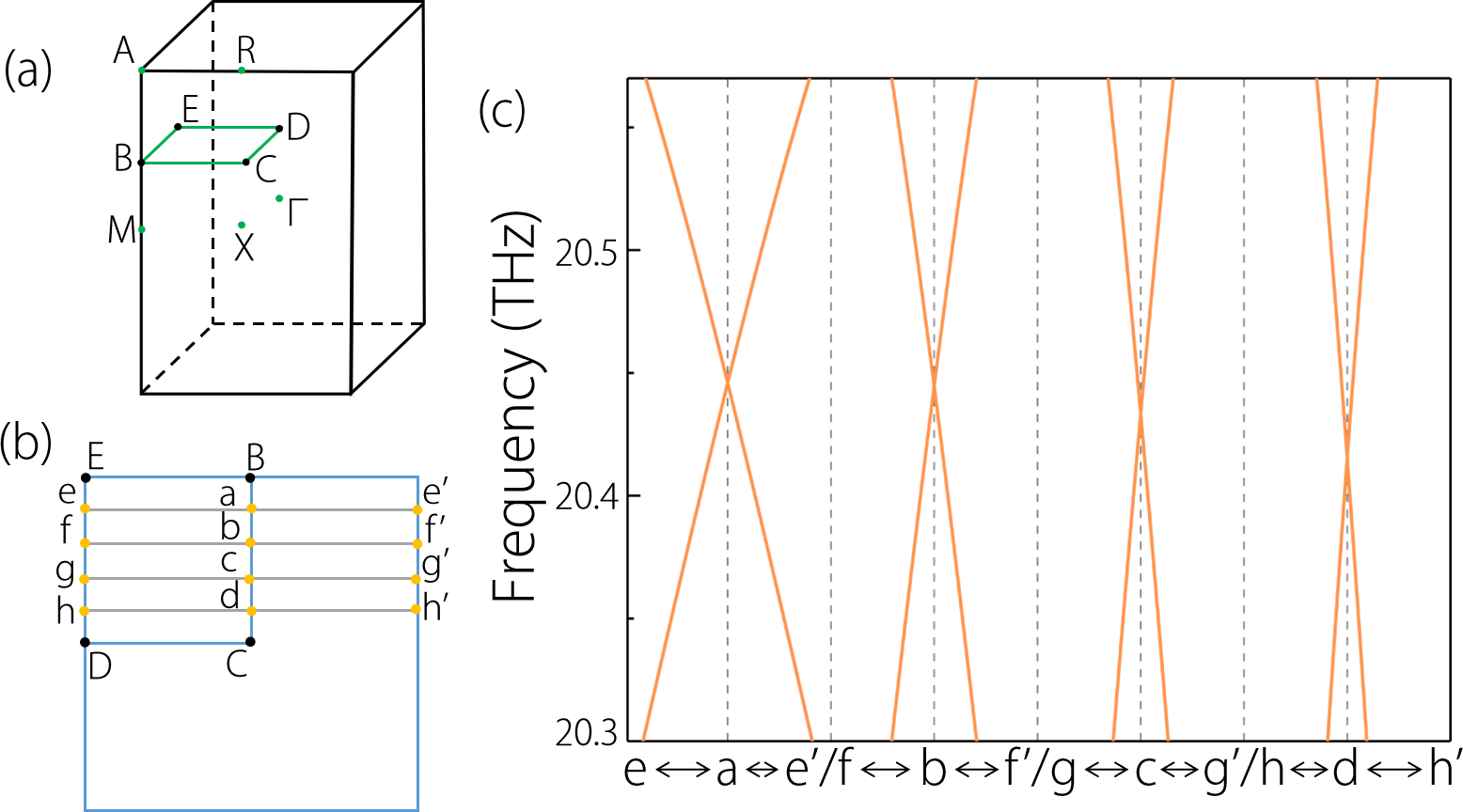}
\caption{(a) The selected B, C, D and E points in the bulk BZ. (b) Some symmetry points of the B--C--D--E plane; a (e), b (f), c (g) and d (h) are located at equal distances between B (E) and C (D). (c) Phonon dispersions along the e--a--e$^{\prime}$, f--b--f$^{\prime}$, g--c--g$^{\prime}$ and h--d--h$^{\prime}$ paths. The a, b, c and d points lie in the $k_y=\pi$ plane.
\label{fig3}}
\end{figure}

In R1, the two nondegenerate phonon bands along the $\Gamma$--X path merged into a two-fold degenerate phonon band along the X--M--A--R--X path, and the phonon bands at the X point exhibit a linear phonon band dispersion. As shown in Fig.~\ref{fig2}(b), the high-symmetry X--M--A--R--X path is present in the $k_y=\pi$ plane of the three-dimensional BZ. In Fig.~\ref{fig3}(b), we selected two symmetry points B and C (E and D) in the $k_y=\pi$ ($k_y=0$) plane and divided the B--C (E--D) path into five parts. Then, some symmetry points along the B--C and E--D paths were chosen. The phonon dispersions along e--a--e$^{\prime}$, f--b--f$^{\prime}$, g--c--g$^{\prime}$ and h--d--h$^{\prime}$ are shown in Fig.~\ref{fig3}(c). Obviously, two nondegenerate phonon bands linearly cross at the a, b, c and d points; the a, b, c and d points lie in the $k_y=\pi$ plane. Actually, the two nondegenerate phonon bands degenerated in the entire $k_y=\pi$ plane and yielded two-dimensional nodal surface phonons~\cite{add57} in the momentum space (see the schematic diagram of nodal surface phonons in Fig.~\ref{fig1}(d)). Similar nodal surface phonons can also be observed in the entire $k_x=\pi$ plane.

The nodal surface in the $k_x=\pi$ plane is due to Kramers degeneracy. Due to the combination of two-fold screw-rotational symmetry $\{C_{2x}|\frac{1}{2}\frac{1}{2}\frac{1}{2}\}$ and time-reversal symmetry $\cal{T}$ (notice that ${\cal{T}}^2 \equiv 1$ for spinless case), ${(\{C_{2x}|\frac{1}{2}\frac{1}{2}\frac{1}{2}\}\cal{T})}^2=-1$ for each point in the $k_x=\pi$ plane; this affords Kramers-like band degeneracy on the $k_x=\pi$ plane and the appearance of the nodal surface state. Similarly, ${(\{C_{2y}|\frac{1}{2}\frac{1}{2}\frac{1}{2}\}\cal{T})}^2=-1$  in the $k_y=\pi$ plane, which guarantees another nodal surface in the $k_y=\pi$ plane~\cite{add15}.

\begin{figure}
\includegraphics[width=8.5cm]{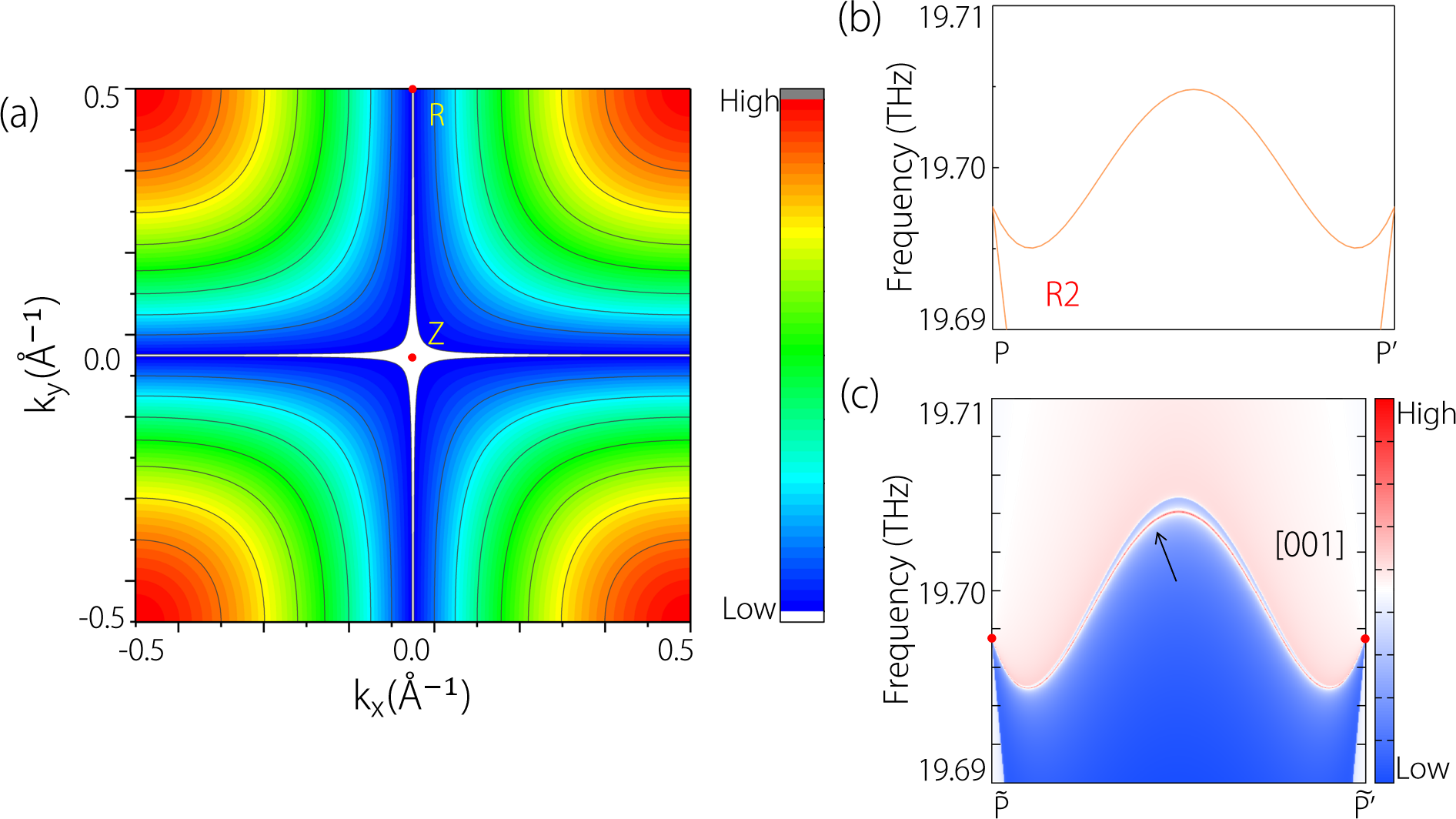}
\caption{(a) The shapes of the WNLs (white lines) in the $k_z=\pi$ plane; (b) the phonon dispersion of SnO$_2$ along the P--$\rm{P}^\prime$ path. (c) Phonon surface spectra along the $\widetilde{\rm{P}}$--$\widetilde{\rm{P}^\prime}$ path on the [001] surface.
\label{fig4}}
\end{figure}

In the R2 region, a two-fold degenerate phonon band exists along the R--Z path. This two-fold degenerate phonon band is present in the $k_z=\pi$ plane and should belong to one pair of Weyl nodal lines (WNLs) in the $k_z=\pi$ plane (see the shapes of the WNLs of the $k_z=\pi$ plane in Fig.~\ref{fig4}(a)). Notice that the R--Z path exhibits a combination of glide-mirror operation and time reversal symmetry $g=\{\sigma_y|\frac{1}{2}\frac{1}{2}\frac{1}{2}\}\cal{T}$, which acts on the lattice momentum as $g:(k_x,k_y,k_z)\rightarrow(-k_x,k_y,-k_z)$ and satisfies $g^2=-1$ in the R--Z path. Additionally, this Kramers-like symmetry guarantees the double band degeneracy on the corresponding line~\cite{add58}. Furthermore, we calculated the band representation of the R--Z high-symmetry line (see Fig~\ref{fig2}(a)). Based on the calculated band representation, the $k \cdot p$ Hamiltonian for the WNL at the R--Z path can be expressed as\begin{equation}\label{1}
\cal{H}_{\textit{WNL}}=\quad\begin{pmatrix}c_1k_y & c_2k_z+ic_3k_x \\ c_2k_z-ic_3k_x & c_1k_y\end{pmatrix},
\end{equation}where $c_i$'s are real parameters.

The Berry phase for a closed loop surrounding the WNLs is calculated based on the following formula:\begin{equation}\label{2}
P_B=\oint_LA(k)\cdot dk,
\end{equation}where $A(k)=-i\langle\varphi(k)|\bigtriangledown_k|\varphi(k)\rangle$ is the Berry connection and $\varphi(k)$ is the periodic part of the Bloch function. For the WNLs in the $k_z=\pi$ plane, $P_B=\pi$ (see Fig.~\ref{fig1}(d)), which implies that the WNLs are topologically nontrivial. Fig.~\ref{fig4}(c) shows the phonon surface spectra along the $\widetilde{\rm{P}}$--$\widetilde{\rm{P}^\prime}$ path on the [001] surface; the positions of the two phonon band-crossings, at P and $\rm{P}^\prime$, of the WNLs are marked by red circles. The phonon dispersion of the bulk along the P--$\rm{P}^\prime$ path is shown in Fig.~\ref{fig4}(b) for a reference. As shown in Fig.~\ref{fig4}(c), the drumhead-shaped phonon surface states (marked by black arrows) arising from the WNLs are obvious.

\begin{figure}
\includegraphics[width=8.5cm]{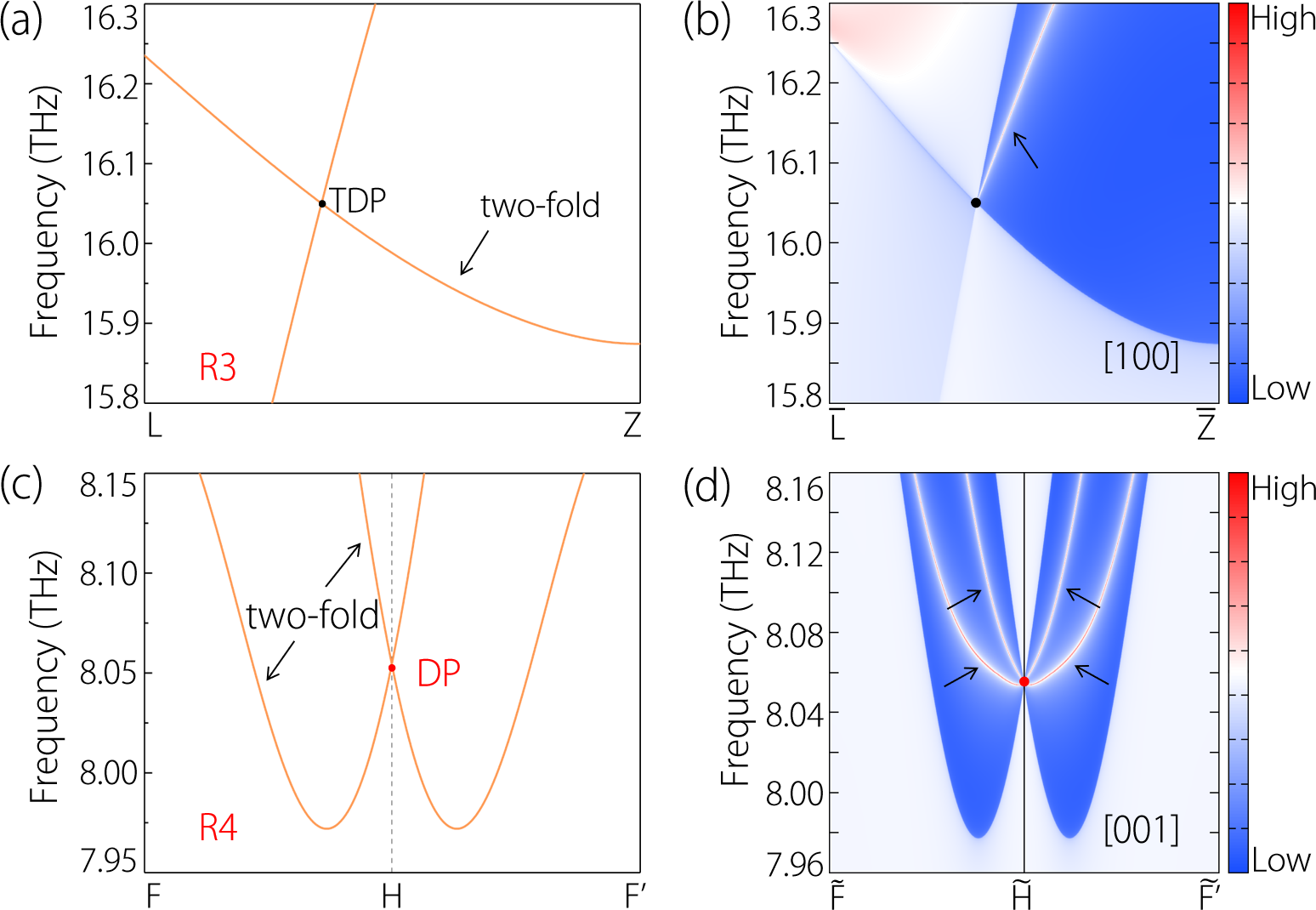}
\caption{(a) and (c) Phonon dispersions along L--Z and F--H--F$^\prime$, respectively; (b) and (d) phonon surface spectra along the $\overline{\rm{L}}$--$\overline{\rm{Z}}$ path on the [100] surface and $\widetilde{\rm{F}}$--$\widetilde{\rm{H}}$--$\widetilde{\rm{F}^\prime}$ path on the [001] surface, respectively. The black circle in (b) and the red circle in (d) represent the positions of TDP and DP, respectively.
\label{fig5}}
\end{figure}
In R3, a phonon band-crossing point with three-fold degeneracy exists along the $\Gamma$--Z path. The three-fold degenerate band-crossing point is actually a triply degenerate point (TDP)~\cite{add41,add59} (see Fig.~\ref{fig5}(a)), which is formed by a two-fold degenerate phonon band and a nondegenerate phonon band. By calculating the band representation of $\Gamma$--Z, we find that TDP is actually an accidental degeneracy formed by a two-dimensional IRR $R_5^\Lambda$ and a one-dimensional IRR $R_1^\Lambda$, as shown in Fig.~\ref{fig2}(a). Using the obtained IRR, the $k \cdot p$ Hamiltonian of the TDP can be directly written as\begin{equation}\label{3}
\cal{H}_{\textit{TDP}=}\quad\begin{pmatrix}c_1k_z & c_2k_y & c_2k_x \\ c_2k_y & c_1k_z & 0 \\ c_2k_x & 0 & c_1k_z\end{pmatrix},
\end{equation}where $c_i$ are real parameters.

Fig.~\ref{fig5}(b) depicts the phonon surface spectra along the $\overline{\rm{L}}$--$\overline{\rm{Z}}$ path on the [100] surface. The phonon surface state around the TDP (black circle) is highlighted with black arrows.

Moreover, in R4, a four-fold degenerate phonon band-crossing point, formed by two two-fold degenerate phonon bands, can be found along the M--A path. Such a four-fold degenerate band-crossing along the M--A path should be a Dirac point (DP)~\cite{add60}. Similarly, we calculate the IRR of the four bands forming the DP in the M--A path. The four bands forming the DP become a pair of two-fold degenerate bands in the M--A path, for which the band representations are $R_7^V R_8^V$ and $R_{10}^V$, as illustrated in Fig.~\ref{fig2}(a). $R_7^V R_8^V$ is a two-dimensional co-representation, where two one-dimensional representations $R_7^V$ and $R_8^V$ are bound together by the anti-unitary operation $\cal{IT}$ and $\cal{I}$ is the inversion symmetry. The corresponding $k\cdot p$ Hamiltonian of DP in the M--A path is
\begin{widetext}
\begin{equation}\label{4}
\cal{H}_{\textit{DP}=}
\quad\begin{pmatrix}ck_z&0&\alpha(k_x+k_y)&-i\alpha(k_x-k_y)\\0&ck_z&\alpha^\ast(k_x+k_y)&i\alpha^\ast(k_x-k_y)\\\alpha^\ast(k_x+k_y)&\alpha(k_x+k_y)&ck_z &0\\ i\alpha^\ast(k_x-k_y)&-i\alpha(k_x-k_y)&0& ck_z\end{pmatrix},
\end{equation}
\end{widetext}where $c$ is a real parameter and $\alpha$ is a complex parameter.

Fig.~\ref{fig5}(d) shows the phonon surface spectra along the $\widetilde{\rm{F}}$--$\widetilde{\rm{H}}$--$\widetilde{\rm{F}^\prime}$ path on the [001] surface (Fig.~\ref{fig1}(d)); the DP position is marked by a red circle. As reference, the phonon dispersion of the bulk along F--H--F$^\prime$ is exhibited in Fig.~\ref{fig5}(c), and DP (at H point) belongs to type \uppercase\expandafter{\romannumeral1} with linear phonon band dispersion. Two phonon surface states (marked with black arrows) originating from DP are visible in Fig.~\ref{fig5}(d).

In summary, we proposed that the already prepared SnO$_2$ with \textit{P}4$_2$/\textit{mnm} structure is a topological phononic material with zero-dimensional TDP and DP phonons, one-dimensional WNL phonons, and two-dimensional nodal surface phonons. TDP and DP are along the $\Gamma$--Z and M--A paths, respectively. One pair of WNLs is in the $k_z=\pi$ plane, and the WNLs are topologically nontrivial due to the $\pi$ Berry phase. Multiple nodal surface states are present in the $k_{x/y}=\pi$ planes.

Before closing, note the following: (i) This study for the first time, to the best of our knowledge, reports that realistic SnO$_2$~\cite{add49} simultaneously hosts zero-dimensional TDP and DP phonons, one-dimensional WNLs phonons and two-dimensional nodal surface phonons. Although some previous studies proved that zero-, one- and two-dimensional degeneracy coexisted in some electronic systems~\cite{add46,add47,add48}, not all the zero-, one- and two-dimensional topological elements were located at or around the Fermi level. Some additional technologies~\cite{add47}, such as hole or electron doping method, should be used to tune the topological elements to the Fermi level. The topological phonons proposed herein do not follow the limitations of the Pauli exclusion principle, and therefore, these topological phonons can be physically observed in the entire range of the phonon frequencies. (ii) The zero-, one- and two-dimensional topological elements coexist in some electronic systems, and their degenerate electronic structures are usually broken by the effect of spin-orbit coupling (SOC)~\cite{add46}. However, the phonons, without SOC, provide us a new platform to search for material samples with true zero-, one- and two-dimensional topological elements in their phonon dispersion. (iii) Topological electronic systems with different topological elements enjoy different physical properties. Therefore, if zero-, one- and two-dimensional topological elements coexisted in one single solid-state material, the study of the entanglement among these three topological elements would be very interesting; (iv) the nodal surface and WNL states shown herein are nearly flat, as shown in Fig.~\ref{fig2}(a). Moreover, as exhibited in Fig.~\ref{fig3}(c), an important feature of the proposed nodal surface phonons is that they host linear phonon band dispersion along the direction normal to the surface~\cite{add15}. (v) The phonon surface states for the WNL, DP and TDP phonons are obvious, benefiting experimental detection. (vi) Finally, since 1961, the preparation technology of high-quality pure single crystals of SnO$_2$ has rapidly developed~\cite{add56}. Till date, more than ten methods~\cite{add56} have been developed to experimentally obtain pure single crystals of SnO$_2$. We believe that SnO$_2$ is a good candidate for investigating the zero-, one- and two-dimensional topological phonons. Thus, the experimental confirmation of the predicted topological elements in SnO$_2$ phonons is imminent.

\emph{\textcolor{blue}{Acknowledgments}} Z. Z. is grateful for the support from the National Natural Science Foundation of China (Grants No. 12004028). X. W. is grateful for the support from the National Natural Science Foundation of China (No. 51801163) and the Natural Science Foundation of Chongqing (No. cstc2018jcyjA0765).


\begin{thebibliography}{}
\bibitem{add1} Burkov, A. A. (2016). Nature Materials, 15(11), 1145-1148.
\bibitem{add2} Narang, P., Garcia, C. A. and Felser, C. (2021). Nature Materials, 20(3), 293-300.
\bibitem{add3} Gao, H., Venderbos, J. W., Kim, Y. and Rappe, A. M. (2019). Annual Review of Materials Research,49, 153-183.
\bibitem{add4} Lv, B. Q., Qian, T. and Ding, H. (2021). Reviews of Modern Physics, 93(2), 025002.
\bibitem{add5} Schoop, L. M., Pielnhofer, F. and Lotsch, B. V. (2018). Chemistry of Materials, 30(10), 3155-3176.
\bibitem{add6} Wang S., Wu W. and Yang S. (2019). Acta Physica Sinica, 68(22), 224206.
\bibitem{add7} Fang, C., Weng, H., Dai, X. and Fang, Z. (2016). Chinese Physics B, 25(11), 117106.
\bibitem{add8} Fang, C., Chen, Y., Kee, H. Y. and Fu, L. (2015). Physical Review B, 92(8), 081201.
\bibitem{add9} Burkov, A. A., Hook, M. D. and Balents, L. (2011). Physical Review B, 84(23), 235126.
\bibitem{add10} Chen, C., Wang, S. S., Liu, L., Yu, Z. M., Sheng, X. L., Chen, Z. and Yang, S. A. (2017). Physical Review Materials, 1(4), 044201.
\bibitem{add11} Zhu, Z., Liu, Y., Yu, Z. M., Wang, S. S., Zhao, Y. X., Feng, Y., \textit{et. al.} (2018). Physical Review B, 98(12), 125104.
\bibitem{add12} Wu, W., Jiao, Y., Li, S., Sheng, X. L., Yu, Z. M. and Yang, S. A. (2019). Physical Review Materials, 3(5), 054203.
\bibitem{add13} Zhang, X., Yu, Z. M., Lu, Y., Sheng, X. L., Yang, H. Y. and Yang, S. A. (2018). Physical Review B, 97(12), 125143.
\bibitem{add14} Li, S., Yu, Z. M., Liu, Y., Guan, S., Wang, S. S., Zhang, X., \textit{et. al.} (2017). Physical Review B, 96(8), 081106.
\bibitem{add15} Wu, W., Liu, Y., Li, S., Zhong, C., Yu, Z. M., Sheng, X. L., \textit{et. al.} (2018). Physical Review B, 97(11), 115125.
\bibitem{add16} Gibson, Q. D., Schoop, L. M., Muechler, L., Xie, L. S., Hirschberger, M., Ong, N. P., \textit{et. al.} (2015). Physical Review B, 91(20), 205128.
\bibitem{add17} Zhang, T., Jiang, Y., Song, Z., Huang, H., He, Y., Fang, Z., \textit{et. al.} (2019). Nature, 566(7745), 475-479.
\bibitem{add18} Song, Z., Zhang, T. and Fang, C. (2018). Physical Review X, 8(3), 031069.
\bibitem{add19} Vergniory, M. G., Elcoro, L., Felser, C., Regnault, N., Bernevig, B. A. and Wang, Z. (2019). Nature, 566(7745), 480-485.
\bibitem{add20} Tang, F., Po, H. C., Vishwanath, A. and Wan, X. (2019). Nature, 566(7745), 486-489.
\bibitem{add21} Kruthoff, J., De Boer, J., Van Wezel, J., Kane, C. L. and Slager, R. J. (2017). Physical Review X, 7(4), 041069.
\bibitem{add22} Po, H. C., Vishwanath, A. and Watanabe, H. (2017). Nature Communications, 8(1), 1-9.
\bibitem{add23} Song, Z., Zhang, T., Fang, Z. and Fang, C. (2018). Nature Communications, 9(1), 1-7.
\bibitem{add24} Bradlyn, B., Elcoro, L., Cano, J., Vergniory, M. G., Wang, Z., Felser, C., \textit{et. al.} (2017). Nature, 547(7663), 298-305.
\bibitem{add25} Yu, Z. M., Zhang, Z., Liu, G. B., Wu, W., Li, X. P., Zhang, R. W., \textit{et. al.} arXiv preprint arXiv:2102.01517.
\bibitem{add26} Miao, H., Zhang, T. T., Wang, L., Meyers, D., Said, A. H., Wang, Y. L., \textit{et. al.} (2018). Physical Review Letters, 121(3), 035302.
\bibitem{add27} Jin, Y., Wang, R. and Xu, H. (2018). Nano Letters, 18(12), 7755-7760.
\bibitem{add28} Wang, R., Xia, B. W., Chen, Z. J., Zheng, B. B., Zhao, Y. J. and Xu, H. (2020). Physical Review Letters, 124(10), 105303.
\bibitem{add29} Jin, Y. J., Chen, Z. J., Xia, B. W., Zhao, Y. J., Wang, R. and Xu, H. (2018). Physical Review B,98(22), 220103.
\bibitem{add30} Zhang, T. T., Miao, H., Wang, Q., Lin, J. Q., Cao, Y., Fabbris, G., \textit{et. al.} (2019). Physical Review Letters, 123(24), 245302.
\bibitem{add31} Liu, Q. B., Fu, H. H., Xu, G., Yu, R. and Wu, R. (2019). The Journal of Physical Chemistry Letters, 10(14), 4045-4050.
\bibitem{add32} Liu, Q. B., Qian, Y., Fu, H. H. and Wang, Z. (2020). npj Computational Materials, 6(1), 1-6.
\bibitem{add33} Li, J., Wang, L., Liu, J., Li, R., Zhang, Z. and Chen, X. Q. (2020). Physical Review B,101(8), 081403.
\bibitem{add34} Liu, J., Hou, W., Wang, E., Zhang, S., Sun, J. T. and Meng, S. (2019). Physical Review B,100(8), 081204.
\bibitem{add35} Li, J., Liu, J., Baronett, S. A., Liu, M., Wang, L., Li, R., \textit{et. al.} (2021). Nature Communications, 12(1), 1-12.
\bibitem{add36} Liu, P. F., Li, J., Tu, X. H., Li, H., Zhang, J., Zhang, P., \textit{et. al.} (2021). Physical Review B, 103(9), 094306.
\bibitem{add37} Zheng, B., Xia, B., Wang, R., Chen, Z., Zhao, J., Zhao, Y. and Xu, H. (2020). Physical Review B, 101(10), 100303.
\bibitem{add38} Liu, Q. B., Wang, Z. and Fu, H. H. (2021). Physical Review B, 103(16), L161303.
\bibitem{add39} Jin, Y. J., Chen, Z. J., Xiao, X. L. and Xu, H. (2021). Physical Review B, 103(10), 104101.
\bibitem{add40} Liu, Y., Chen, X. and Xu, Y. (2020). Advanced Functional Materials, 30(8), 1904784.
\bibitem{add41} Singh, S., Wu, Q., Yue, C., Romero, A. H. and Soluyanov, A. A. (2018). Physical Review Materials, 2(11), 114204.
\bibitem{add42} Li, N., Ren, J., Wang, L., Zhang, G., H$\ddot{\rm{a}}$nggi, P. and Li, B. (2012). Reviews of Modern Physics, 84(3), 1045.
\bibitem{add43} Huber, S. D. (2016). Nature Physics, 12(7), 621-623.
\bibitem{add44} Liu, Y., Xu, Y. and Duan, W. (2018). National Science Review, 5(3), 314-316.
\bibitem{add45} Long, Y., Ren, J. and Chen, H. (2020). Physical Review Letters, 124(18), 185501.
\bibitem{add46} Yang, T., Liu, Y., Wu, Z., Wang, X. and Zhang, G. (2021). Materials Today Physics, doi: 10.1016/j.mtphys.2021.100348.
\bibitem{add47} Wang, X., Cheng, Z., Zhang, G., Wang, B., Wang, X. L. and Chen, H. (2020). Nanoscale, 12(15), 8314-8319.
\bibitem{add48} Yang, T., Ding, G., Cheng, Z., Wang, X. and Zhang, G. (2020). Journal of Materials Chemistry C, 8(23), 7741-7748.
\bibitem{add49} Baur, W. H. (1956). Acta Crystallographica,9(6), 515-520.
\bibitem{add50} Perdew, J. P., Burke, K. and Ernzerhof, M. (1998). Physical Review Letters, 80(4), 891.
\bibitem{add51} Bl$\ddot{\rm{o}}$chl, P. E. (1994). Physical review B, 50(24), 17953.
\bibitem{add52} Togo, A. and Tanaka, I. (2015). Scripta Materialia, 108, 1-5.
\bibitem{add53} Wu, Q., Zhang, S., Song, H. F., Troyer, M. and Soluyanov, A. A. (2018). Computer Physics Communications, 224, 405-416.
\bibitem{add54} Liu, G. B., Chu, M., Zhang, Z., Yu, Z. M. and Yao, Y. (2021). Computer Physics Communications, 107993.
\bibitem{add55} Bradley, C. and Cracknell, A. (2009). The mathematical theory of symmetry in solids: representation theory for point groups and space groups. Oxford University Press.
\bibitem{add56} Jarzebski, Z. M. and Marton, J. P. (1976). Journal of the electrochemical Society, 123(7), 199C.
\bibitem{add57} Wang, X., Zhou, F., Yang, T., Kuang, M., Yu, Z. M. and Zhang, G. (2021). arXiv preprint arXiv:2103.15495.
\bibitem{add58} Young, S. M. and Kane, C. L. (2015). Physical Review Letters, 115(12), 126803.
\bibitem{add59} Winkler, G. W., Singh, S. and Soluyanov, A. A. (2019). Chinese Physics B, 28(7), 077303.
\bibitem{add60} Chen, Z. J., Wang, R., Xia, B. W., Zheng, B. B., Jin, Y. J., Zhao, Y. J. and Xu, H. (2021). Physical Review Letters, 126(18), 185301.


\end{thebibliography}
\end{document}